\documentclass[12pt]{article}
\usepackage{epsfig}

\def\epsfg#1#2{\epsfig{file=#1.eps,width=#2}}

\def\i{\mathrm{i}}
\def\e{\mathrm{e}}
\def\d{\mathrm{d}}
\def\half{{\textstyle{1\over2}}}

\title{Soliton Formation in Chiral Quark Models\thanks{%
Talk given at the Mini-Workshop ``Hadrons as Solitons'',
Bled (Slovenia), July 9---17, 1999.}}
\author{Bojan Golli\thanks{%
E-mail: Bojan.Golli@ijs.si}
\\
Faculty of Education, University of Ljubljana and
\\ 
J.~Stefan Institute, Jamova 39, P.O.~Box 3000, 1001 Ljubljana, Slovenia}

\date{}

\begin{document}

\maketitle

\begin{abstract}
We describe how the non-local regularization can be implemented 
in the calculation of solitons in the Nambu Jona-Lasinio model 
as well as in the equivalent linear $\sigma$-model. 
We investigate different forms of regulators and show that the 
3-momentum cut-off leads to serious conceptual difficulties.
\end{abstract}
   
\section{Motivation}

This work was done together with Georges Ripka and Wojciech Broniowski.

Solitons corresponding to baryons have been found in several chiral
quark models. Many of these solutions turn out to be unstable against 
collapse unless additional constraints are introduced in the model.
The well known examples are the linear NJL model with proper time 
regularization \cite{Goeke92,Ripka93d} and the linear $\sigma$-model 
with sea quarks \cite{Ripka87,Perry87}.
Even in the linear $\sigma$-model with only valence quarks the energy 
of the soliton becomes too low for any choice of model parameters
if one goes beyond the mean field approximation.
In all these models the instability occurs because it is energetically 
favorable for the chiral field to acquire arbitrary (or very) high 
gradients.
This suggests that cutting off high momenta in the interaction 
may prevent the collapse and stabilize the soliton.
A simple sharp cut-off does not yield a stable solution 
while a smooth  behavior of the regulator (usually interpreted as 
a $k$-dependent quark mass) can indeed lead to  solitons which are 
stable against the decay into free quarks as well as against collapse.
Such a regularization has a physical justification in 
QCD calculations of the quark propagation in an instanton liquid 
which predict a non-local effective  interaction between quarks 
with a 4-momentum cut-off $\Lambda\sim 600$~MeV \cite{Diakonov86}.

Further physical implications of the non-local regularization 
are discussed in the contributions to this workshop by George Ripka
and Wojciech Broniowski~\cite{bled99}.

\section{The NJL model with non-local regulators}

The non-local regularization of the quark-quark interaction can be 
implemented in the NJL type models by replacing the contact term 
$\left(\bar{q}(x)\Gamma_aq(x)\right)^2$, 
$\Gamma_a\equiv(1,\i\gamma_5\tau_a)$
by a non-local form.
Usually one introduces a regulator $r$ diagonal in 4-momentum 
space such that $q(x) \to \int\d_4y\langle x|r|y\rangle q(y)$.
The QCD derivation of the quark propagation in a dilute instanton 
gas predicts the following functional dependence for 
$r(k^2) =\langle k'|r|k\rangle\delta(k-k')$ \cite{Diakonov86}:
\begin{equation}
 r =  f\left(z\right) =
 -z\frac \d{\d z}\left(I_0\left(z\right) K_0\left(z\right)
                   -I_1\left(z\right) K_1\left(z\right) \right)\;,
\qquad
z = {\sqrt{k^2}\,\rho\over 2}\;,  
\label{instanton}
\end{equation}
where $\rho$ is the instanton size of the order $(600~\rm{MeV})^{-1}$.
As we shall see in the following it is necessary to analytically 
continue the regulator to negative $k^2$ in order to be able to 
treat the valence orbit.
This is not possible with the form (\ref{instanton}) since it has
a cut along the negative real axis starting at $k^2=0$.
We use instead a Gaussian shape of the regulator:
\begin{equation}
  r(k^2) = \e^{-{k^2\over2\Lambda^2}}\;,
\label{gaussian}
\end{equation}
or a ``monopole'' shape:
\begin{equation}
  r(k^2) = {1\over 1 + {k^2\over\Lambda^2}}\;,
\label{pole}
\end{equation}
which has the proper behavior for large $k^2$
where one gluon exchange dominates.

The expression for the energy of the soliton and the self-consistency 
equations can be derived from the bosonized Euclidean action
 \begin{equation}
I=-{\rm Tr}\log \left( -\i\partial_\mu \gamma_\mu 
  +m+r\left(S +\i\gamma_5P_a\tau_a \right)r\right) +
   \frac1{2G^2}\int \d_4x\left( S^2+P_a^2\right)\;,
\label{action}  
\end{equation}
where $S$ and  $P_a$ are the chiral fields and
are the dynamical variables of the system.

The main difficulty is the presence of time in the regulator.
In order to evaluate the trace in (\ref{action}) it is convenient
to introduce energy dependent basis states, which are
solutions of the Dirac equation:
\begin{equation}
  h(\nu^2)\,|q_{j\nu}\rangle=e_j(\nu^2)\, | q_{j\nu}\rangle
\label{Diraceq} 
\end{equation}
with
\begin{equation}
  h\left(\nu^2\right) = -\i\vec \alpha \cdot \nabla 
    + \beta r\!\left(\nu^2-\vec\nabla^2\right) 
     (S(\vec{r}) + \i\gamma_5\tau_aP_a(\vec{r}))
  \,r\!\left(\nu^2-\vec\nabla^2\right) + \beta m\;.
\label{DH}
\end{equation}

From (\ref{action}) the following expression for a stationary
configuration can be derived \cite{GBR98}:
\begin{equation}
  E =  \frac 1{2\pi}\int_{-\infty}^\infty \nu \d\nu
  \sum_j\frac{\i+\frac{\d e_j}{\d\nu}}{\i\nu + e_j\left(\nu^2\right)}
  + \frac 1{2G^2}\int \d_3x\left(S^2 + P_a^2\right) -{\rm vacuum}\;.
\label{energy}
\end{equation}
Note that when no regulator (i.e. $r\equiv 1$) or a time-independent 
regulator is used, the energies $e_j$ are independent of $\nu$ and the 
integration can be carried out using the Cauchy theorem.
Closing the contour from below yields the well known expression 
for the energy of the Dirac sea: $E^{\rm sea}=\sum_{e_j<0}e_j$.
(Note that the energies of occupied orbits lie on the
negative imaginary $\nu$-axis.)

When the soliton describes a baryon, the energy of three valence 
quarks is added to the energy of the Dirac sea.\footnote{%
Only if it is positive, otherwise it is already contained
in the above sum.}
The same result can be formally obtained  by deforming the 
contour in (\ref{energy}) in such a way as to encircle the
valence orbit (for detailed  discussion on this point see
Wojciech Broniowski contribution to this workshop).
Such a prescription gives the expected result
provided the orbits do not depend on $\nu$. 
However,
when the regulator depends on time (or $\nu$), this may not 
lead to the correct result since the regulator generates additional 
poles scattered in the whole complex $\nu$-plane.
It may still work well for an isolated pole on the positive imaginary 
axis close to 0 as is the case of the $0^+$ orbit in the soliton 
with the hedgehog form of the background chiral field \cite{Ripka84}.
This pole can then be treated separately, yielding the valence
contribution to the soliton energy $E^\mathrm{val}=3e_\mathrm{val}$, 
where the energy of the valence orbit is determined from
\begin{equation}
  \left.\i\nu + e_{0^+}(\nu^2)\right|_{\nu^2=-e_\mathrm{val}^2}=0\;.
\label{eval} 
\end{equation}
The soliton energy can now be written as:
\begin{equation}
E_{{\rm sol}} = E^\mathrm{val} +  E^\mathrm{sea} + E^\mathrm{meson}\;.
\label{sole} 
\end{equation}
The sea contribution is
\begin{equation}
   E^{\rm sea} = -N_c\sum_{j\in{\rm all}} \int_0^\infty 
      {\d\nu\over\pi} \,\Biggl[
    {e_j(\nu^2)(e_j(\nu^2)-2\nu^2b_j(\nu^2))\over\nu^2+e_j(\nu^2)^2}
   - {e^0_j(\nu^2)(e^0_j(\nu^2)-2\nu^2b^0_j(\nu^2))
         \over\nu^2+e^0_j(\nu^2)^2}\Biggr]
\end{equation}
with 
$
   b_j(\nu^2) = \partial e_j(\nu^2)/\partial\nu^2
$
and is evaluated by direct numerical integration along the 
real $\nu$-axis.
The term $E^\mathrm{meson}$ is given by the last integral in 
(\ref{energy}) (with the integrand $S^2+P_a^2-M_0^2$).

The above prescription is further supported by the fact
that it gives an exact result for the baryon number, 
which can be expressed as \cite{GBR98}: 
\begin{equation}
  B=-\frac 1{2\pi\i N_c}\int_{-\infty }^\infty \d\nu \,
  \sum_j\frac{\i+\frac{\d e_j(\nu)}{\d\nu }}{\i\nu+e_j(\nu)}\;.  
\label{number}
\end{equation}

The self-consistent equations derived from (\ref{action})
take the form (the hedgehog ansatz, 
$
  P_a(r) = \widehat{r}_a P(r)
$, 
for the pion field is assumed):
\begin{eqnarray}
 \left\{{S(r)\atop P(r)}\right\}
    &=& 
    G^2\Biggl[N_q\,{\rm res}_v^{-1}\tilde{q}_{v_{\nu0}}^\dagger(\vec{r})
     \left\{{\beta\atop\i\beta\gamma_5\tau_a\hat{r}_a}\right\}
     \tilde{q}_{v_{\nu0}}(\vec{r})\Biggr.
\nonumber\\
&& + \Biggl.
    N_c \int_0^\infty {\d\nu\over\pi}\,
    \sum_j{e_j(\nu^2)\over\nu^2+e_j(\nu^2)^2}
     \,\tilde{q}_{j_\nu}^\dagger(\vec{r})
     \left\{{\beta\atop\i\beta\gamma_5\tau_a\hat{r}_a}\right\}
     \tilde{q}_{j_\nu}(\vec{r})\Biggr]\;,
\label{sceq}
\end{eqnarray}
where 
$
  \tilde{q}_{j_\nu}(\vec{r}) = 
     r((\nu^2-\vec{\nabla}^2)/\Lambda^2) q_{j_\nu}(\vec{r})
$
and
$
{\rm res}_v = 1 - \i{\d e_{v_{\nu0}}\over\d\nu}
$
is the residue of the valence pole.

\begin{figure}[t]
\begin{center}
\hrule height 0pt depth 0pt
\epsfg{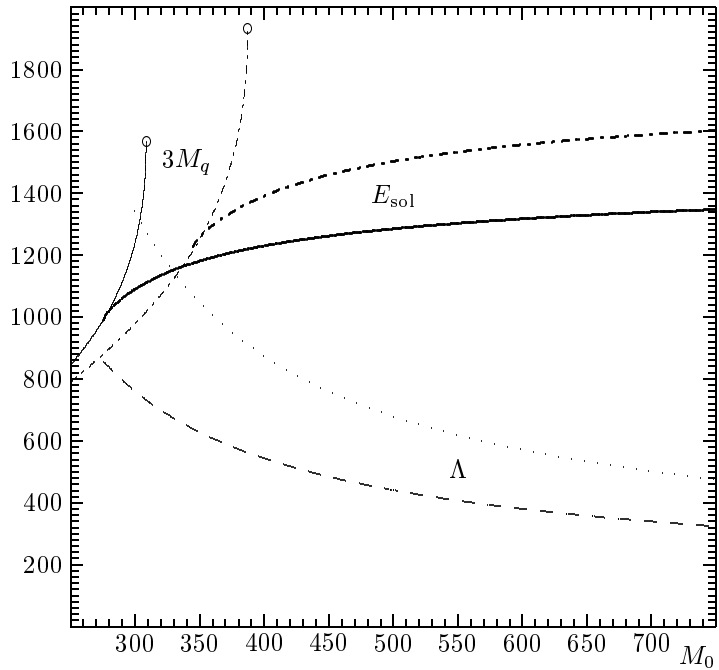}{95mm}
\caption{The energy (in MeV) of the soliton ($E_{\rm sol}$),
3 times the free-space quark mass ($3M_q$) and the cut-off $\Lambda$ 
plotted as functions of the parameter $M_0$ (in MeV).
Two parameterizations of the Gaussian regulator are compared;
(i) solid and dashed lines: $\Lambda$ is fitted to $f_\pi=93$~MeV,
(ii) lines containing dots: $\Lambda$ is fitted to 
$f_\pi =1.25\times 93$~MeV.
For the first parameterization $E_{\rm sol}<3M_q$ at $M_0=276$~MeV
and the soliton becomes stable; 
free quarks do not exist beyond $M_0=309$~MeV. 
For the second parameterization the corresponding values of $M_0$
are $345$~MeV and $387$~MeV.}
\label{figep}
\end{center}
\end{figure}

A necessary condition for a stable soliton configuration is that the
energy (\ref{energy}) is lower than the energy of three free quarks.
When the regulator depends on time, the
free quark mass, $M_q$, is not simply the vacuum value of the
chiral field, $M_0$, but is determined by the position of
the pole of the quark propagator in the vacuum \cite{Ripka97}, 
{\em i.e.} it corresponds to the solution of  
$
   \left. k^2+(r(k^2)^2M_0+m)^2\right|_{k^2=-M_q^2}=0
$.
The solution for real $k^2$ exists only below a critical value of 
$M_0$ (see Figure~\ref{figep}); above this point no stable free 
quarks exist. 
However, a stable solution can always be found beyond this point 
provided the quarks dress in a spatially non-uniform background 
chiral field.

\section{Dependence on the shape of the regulator}

The model (\ref{action}) possesses 4 parameters: the vacuum value
of the chiral field $M_0$, the current quark mass $m$,
the coupling constant $G$ and the cut-off parameter $\Lambda$.
The coupling constant $G$ is fixed from the stationarity condition 
in the vacuum, $\Lambda$ is fitted to the pion decay constant $f_\pi$, 
and $m$ to $m_\pi$. We are left with one free parameter $M_0$.

In this section we analyze how different shapes of the regulator
affect the result.
We compare the Gaussian (\ref{gaussian}), the monopole (\ref{pole}) 
and a modified version of the instanton (\ref{instanton}) regulator.
As we have mentioned it is not possible to continue (\ref{instanton})
to negative $k^2$, which is needed in order to obtain the
valence contribution.
We have instead introduced a form which is identical to the
instanton regulator for $k^2>0$ while its behavior for
$k^2<0$ is replaced by a real function which approximately
follows the real part of (\ref{instanton}) for small negative 
$k^2$ (``extended instanton'').     

The Gaussian and the monopole shapes lead to practically
identical results.
We have also tried other shapes and found very similar results.
The reason is that several properties (including the integral
that determines $f_\pi$) depend mostly on the behavior of the 
regulator for small value of $k^2$.
If this behavior is $r(k^2)\approx 1 - ak^2 + \ldots$
then $a$ is almost uniquely determined by the value of $f_\pi$.
Hence, all shapes with this type of behavior 
lead to very similar results.
The situation is quite different if $\d r(k^2)/\d k^2=0$, 
{\em e.g.\/} for regulators that depend only on $k^4$.
We do not find stable solutions for such regulators;
the energy of the Dirac sea is always higher than the gain due
to the lowering of the valence energy.

The form (\ref{instanton}) has neither of the above behaviors for
$k^2\sim 0$ but behaves as 
$
  1+{3\over16}\,{k^2\over\Lambda^2}\ln{k^2\over\Lambda^2}
$.
It is therefore interesting to check whether 
stable solitons can also be obtained for such a particular form.
We indeed find solitons with very similar properties
to those obtained using (\ref{gaussian}) or (\ref{pole}).

Since $f_\pi$ sets the scale, it is also interesting to study
how a higher value for $f_\pi$ would affect the results.

\def\bstrut{\vrule height 14 pt depth 3 pt width 0pt}

\begin{table}[htp]
\centering
\begin{tabular}{|l|c|c|c|c|c|c|c|c|c|}
\hline

\hline
\bstrut
Regulator & $M_0$ & $\Lambda$ & $m$ & $\langle\bar{q}q\rangle^{1/3}$ 
 & $e_{{\rm val}}$ & $E_{{\rm sea}}$ & 
$E_{{\rm sol}}$ & $\langle r^2\rangle^{1/2}$ & $g_A$ \\
 & [MeV] & [MeV] & [MeV] & [MeV] & [MeV] & [MeV] & [MeV] & fm &  \\ 
\hline
\bstrut 
Gaussian& 350 & 627 & 10.4 & $-200$ & 280 & 1715 & 1180 & 1.04 & 1.16 \\
\bstrut 
Gaussian& 450 & 484 & 15.9 & $-174$ & 266 & 1275 & 1261 & 0.96 & 1.12 \\ 
\hline
\bstrut 
monopole & 350 & 834 & 5.24 & $-252$ & 275 & 2201 & 1176 & 1.05 & 1.30 \\
\bstrut 
monopole & 450 & 639 & 7.56 & $-223$ & 260 & 1628 & 1261 & 0.98 & 1.28 \\ 
\hline
\bstrut 
extanton & 350 & 611 & 4.77 & $-260$ & 300 & 2374 & 1189 & 1.05 & 1.04 \\
\hline
\bstrut 
Gaussian$^*$ & 450 & 759 & 8.75 & $-246$ & 336 & 2121 & 1458 & 0.83 & 1.14 \\
\hline

\hline 
\end{tabular}
\caption{Properties of the self-consistent soliton solutions
for different shapes of the regulator,
``extanton'' stands for the extended instanton regulator with 
$\Lambda=1/\rho$,
Gaussian$^*$ means a Gaussian regulator with $\Lambda$
fitted to  $f_\pi=1.25\times 93$~MeV,
$\langle\bar{q}q\rangle$ is the one-flavor quark condensate.}
\label{table}
\end{table}

\begin{figure}[t]
\begin{minipage}[t]{65mm}
\hrule height 0pt depth 0pt
\epsfg{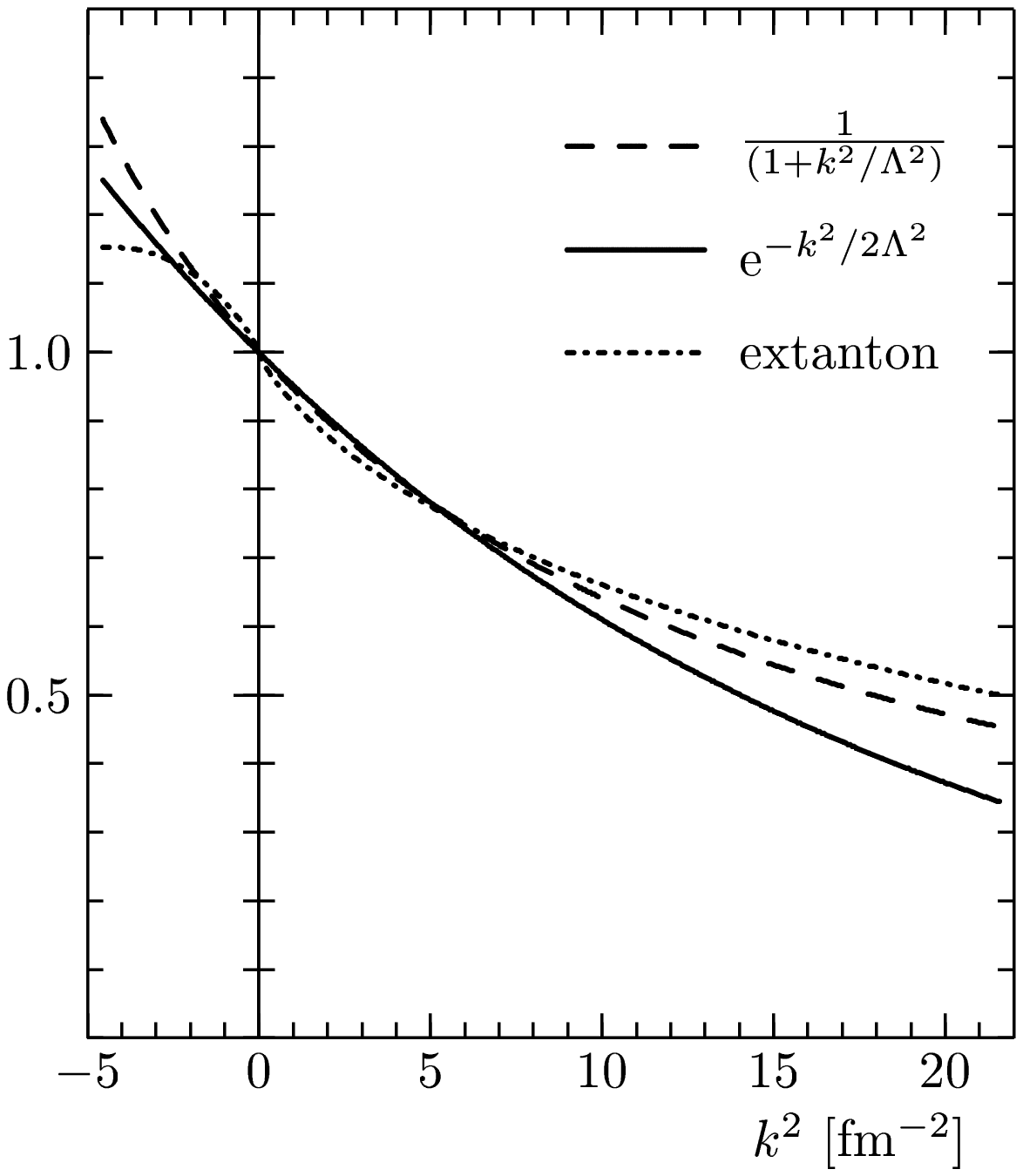}{65mm}
\hrule height 0pt depth 0pt\vspace{-3pt}
\centerline{a)}
\end{minipage}
\hfill
\begin{minipage}[t]{95mm}
\hrule height 0pt depth 0pt
\epsfg{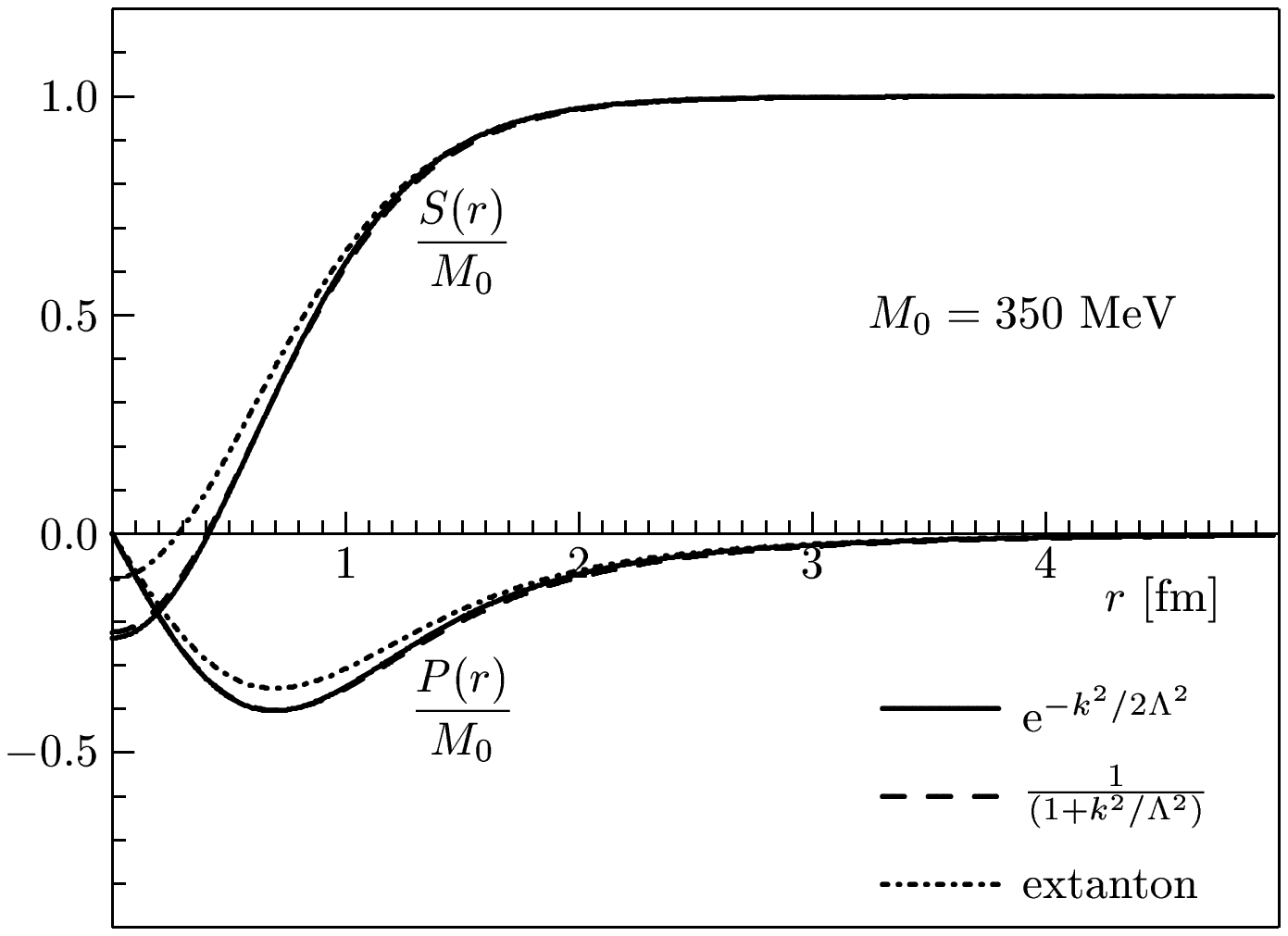}{95mm}
\hrule height 0pt depth 0pt\vspace{15pt}
\centerline{b)}
\end{minipage}
\label{com}
\caption{a) Different shapes of regulators for 
$M_0=350$~MeV and  $\Lambda$ fitted to $f_\pi=93$~MeV,
``extanton'' stands for the extended instanton regulator with 
$\Lambda=1/\rho$.
b) Comparison of the self-consistently determined chiral
field for the three regulators displayed in a).}
\end{figure}

\section{Regulators with a 3-momentum cut-off}

The calculation can be made much simpler and the problem 
of analytic continuation to negative $k^2$ avoided
if we take a regulator that does not depend on time.
Then the states do not depend on $\nu$ and the integration over 
$\nu$  in the above expressions can be carried out analytically.
All quantities still remain finite and the soliton remains stable
against collapse.
The parameters used are the same as for the 4-momentum regulator.
However, the explicit breaking of Lorentz invariance leads 
to serious problems which we discuss in this section.

The energy of the soliton reduces to 
\begin{equation}
E_{{\rm sol}} = N_q e_\mathrm{val} 
                + \sum_{e_j<0}\left(e_j-e^0_j\right) 
                + \frac1{2G^2}\int \d_4x\left(S^2+P_a^2-M_0^2\right)
\label{sole3D} 
\end{equation}
and the self-consistency equation to
\begin{eqnarray}
 \left\{{S(r)\atop P(r)}\right\}
    &=&
      G^2\Biggl[N_q\,q_\mathrm{val}^\dagger(\vec{r})
     \left\{{\beta\atop\i\beta\gamma_5\tau_a\hat{r}_a}\right\}
     r^2((-\vec{\nabla}^2) /\Lambda^2) q_\mathrm{val}(\vec{r})\Biggr. 
\nonumber\\
&& + \Biggl.
    N_c \sum_{e_j<0}\,q_j^\dagger(\vec{r})
     \left\{{\beta\atop\i\beta\gamma_5\tau_a\hat{r}_a}\right\}
     r^2((-\vec{\nabla}^2) /\Lambda^2) q_j(\vec{r})\Biggr]\;.
\nonumber
\end{eqnarray}

The simplest form of the regulator is a  sharp cut-off
limiting the 3-momenta to $\vec{k}^2<\Lambda^2$.
Since one usually uses basis states with good $|\vec{k}|$, 
this is equivalent to restricting the Hilbert space.
Such a form is very easy to implement since the regulator 
does not appear explicitly in the calculation.
Unfortunately, no stable soliton solution is found in this
case; the contribution of the sea quarks is always larger 
than the gain due to the lowering of the valence energy.

\begin{figure}[t]
\begin{center}
\hrule height 0pt depth 0pt
\epsfg{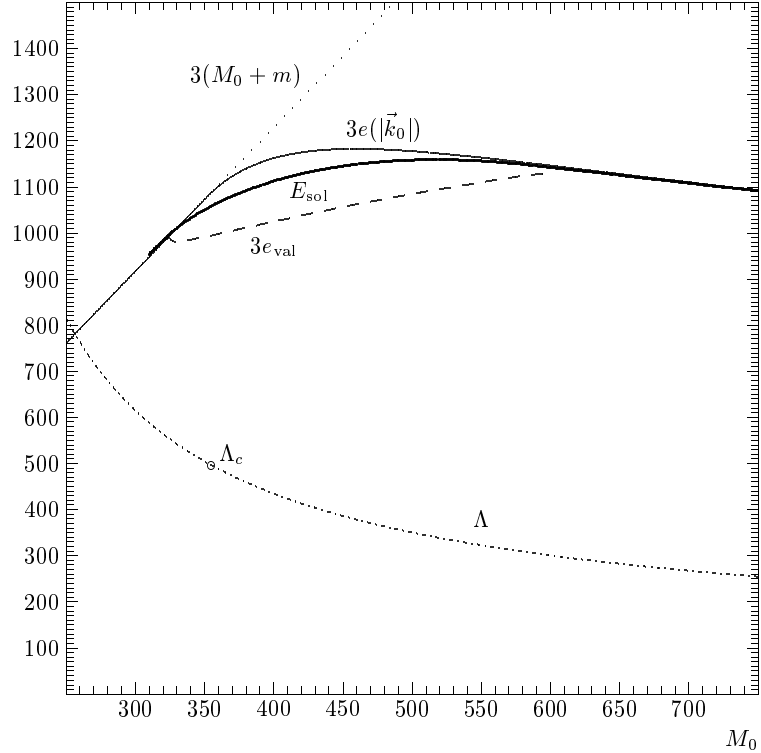}{100mm}
\caption{Solutions obtained with a 3-momentum cut-off: 
the energy (in MeV) of the soliton 
(bold solid line), 3 times the free quark mass (solid line) 
and the valence contribution to the soliton energy (dashed line) are
plotted as functions of $M_0$ (in MeV).
The energy of the lowest free quark state is equal to $M_0+m$ only below
$M_0=355$~MeV, above this value there exist states with $|\vec{k}|>0$ 
which have lower energies.
The value $|\vec{k_0}|$ corresponds to the lowest free quark energy. 
A Gaussian form is used for the regulator, the cut-off $\Lambda$
(dots) is fitted to $f_\pi=93$~MeV.
}
\label{fige3D}
\end{center}
\end{figure}

Stable solutions can be obtained if one takes a smooth
form of the regulator, e.g. a Gaussian form 
$r(\vec{k}^2) = \e^{-\vec{k}^2/2\Lambda^2}$.
The results are displayed in Figure~\ref{fige3D}.
The threshold value of $M_0$ below which the soliton does 
not exist is somewhat larger than for the 4-momentum 
Gaussian regulator.
One notices a rather striking feature that the soliton
does not exist {\em beyond\/} a certain value of $M_0$.
The reason for such behavior is the following:
the energy of a free quark can be written as 
$
  e(|\vec{k}|) = 
         \sqrt{\vec{k}^2 + M_0^2r^4(\vec{k}^2/\Lambda^2)}
$.
For sufficiently small $\Lambda/M_0$ the minimum of $e(|\vec{k}|)$ 
is not at $|\vec{k}|=0$ but rather at some $|\vec{k_0}|>0$,
and the energy of a free quark becomes smaller than $M_0$.
Increasing $M_0$ the cut-off $\Lambda$ decreases and at a certain 
value it becomes energetically favorable for the quarks in the 
soliton to acquire sufficiently high momenta and leave the soliton.
This happens when $3e(|\vec{k_0}|) < E_{{\rm soliton}}$.
Such an unphysical behavior is a clear consequence of
breaking the Lorentz invariance in the interaction.

The value of $|\vec{k_0}|$ can be easily determined if we choose 
a Gaussian regulator and $m=0$.
For the values of $\Lambda$ below 
$\Lambda_c=\sqrt{2}M_0$ the minimum of
$
  e(|\vec{k}|) = \sqrt{\vec{k}^2 + M_0^2\e^{-2\vec{k}^2/\Lambda^2}}
$
occurs for 
$
   \vec{k}_0^2 = \half\Lambda^2\ln({{2M_0^2/\Lambda^2}})
$
and the free quark energy at the minimum is:
$
  e(|\vec{k_0}|) = \Lambda\sqrt{{\half(1+\ln({{2M_0^2/\Lambda^2}}))}}
$. 
For $m\ne 0$ this value is slightly modified:
$$
  e(|\vec{k_0}|) = 
     \Lambda\sqrt{{f_c+\ln{{2f_cM_0^2\over\Lambda^2}}\over2}}\;,
\qquad
  f_c = \left[\sqrt{1 + {m^2\over2\Lambda^2}} 
              + {m\over\sqrt{2}\Lambda}\right]^2\;.
$$ 

Figure~\ref{fige3D} shows that the soliton exists 
for $325~{\rm MeV}<M_0<600$~MeV.
At $M_0=355$~MeV the cut-off $\Lambda$ reaches the critical value
and the lowest free quark state has $|\vec{k}|>0$.
Nonetheless, the soliton remains bound since 
$3e(|\vec{k_0}|)>E_{{\rm soliton}}$.
At $M_0\approx  600$~MeV the condition is no longer fulfilled 
and no bound solution exists beyond this point.
Between these two values the energy of the soliton  differs only
little from the energy of three free quarks. 
The soliton is weakly bound;
the chiral field stays close to its vacuum value while
the soliton radius is large.

\section{Solitons in the equivalent linear $\sigma$-model}

It is well known that the NJL model can be transformed 
into the form of an equivalent linear $\sigma$-model.
This transformation is explained in \cite{Ripka97}, chapter 5.
If one assumes a sharp cut-off and sufficiently large $\Lambda$
(compared to $M_0$)  the Lagrangian density of the equivalent 
$\sigma$-model takes the familiar form
\begin{equation}
 \mathcal{L}_\sigma = \half(\partial_\mu\sigma)^2 +
                      \half(\partial_\mu\pi_a)^2 
    - {\lambda^2\over 4}\left(\sigma^2+\pi_a^2-f_\pi^2\right)^2
    - \half m_\pi^2\left((\sigma-f_\pi)^2+\pi_a^2\right)\;. 
\label{elsm-fam}
\end{equation}
The fields $\sigma$ and $\pi_a$ are related to the two components
of the chiral field of the bosonized NJL model as
$
  \sigma(x) = S(x)/g
$,
$
  \pi_a(x) = P_a(x)/g
$,
where
$
   g = {M_0/f_\pi}
$.
The parameter $\lambda$ and the mass of the $\sigma$ meson 
are related to the parameters of the NJL model by\footnote{%
The $\lambda^2$ here is a factor 2 smaller than the one in 
\cite{Ripka97}.}
$
   g = f_\pi/M_0
 $,
$
   \lambda^2 = 2g^2
$
and
$
   m_\sigma^2 = 4M_0^2 + m_\pi^2\;. 
$
The question remains whether the above assumptions are met in
the model described in section 2.\footnote{%
In fact, they are not; the value of $M_0/\Lambda$ is close to 1
while the smooth form of the regulator may and does generate
additional terms in the Lagrangian (\ref{elsm-fam}) and modifies
the values of the parameters.
A work that will take into account the additional terms is 
in progress.
The conclusions in this section remain valid but the results
should be considered only as qualitative.}

\begin{figure}[t]
\begin{center}
\epsfg{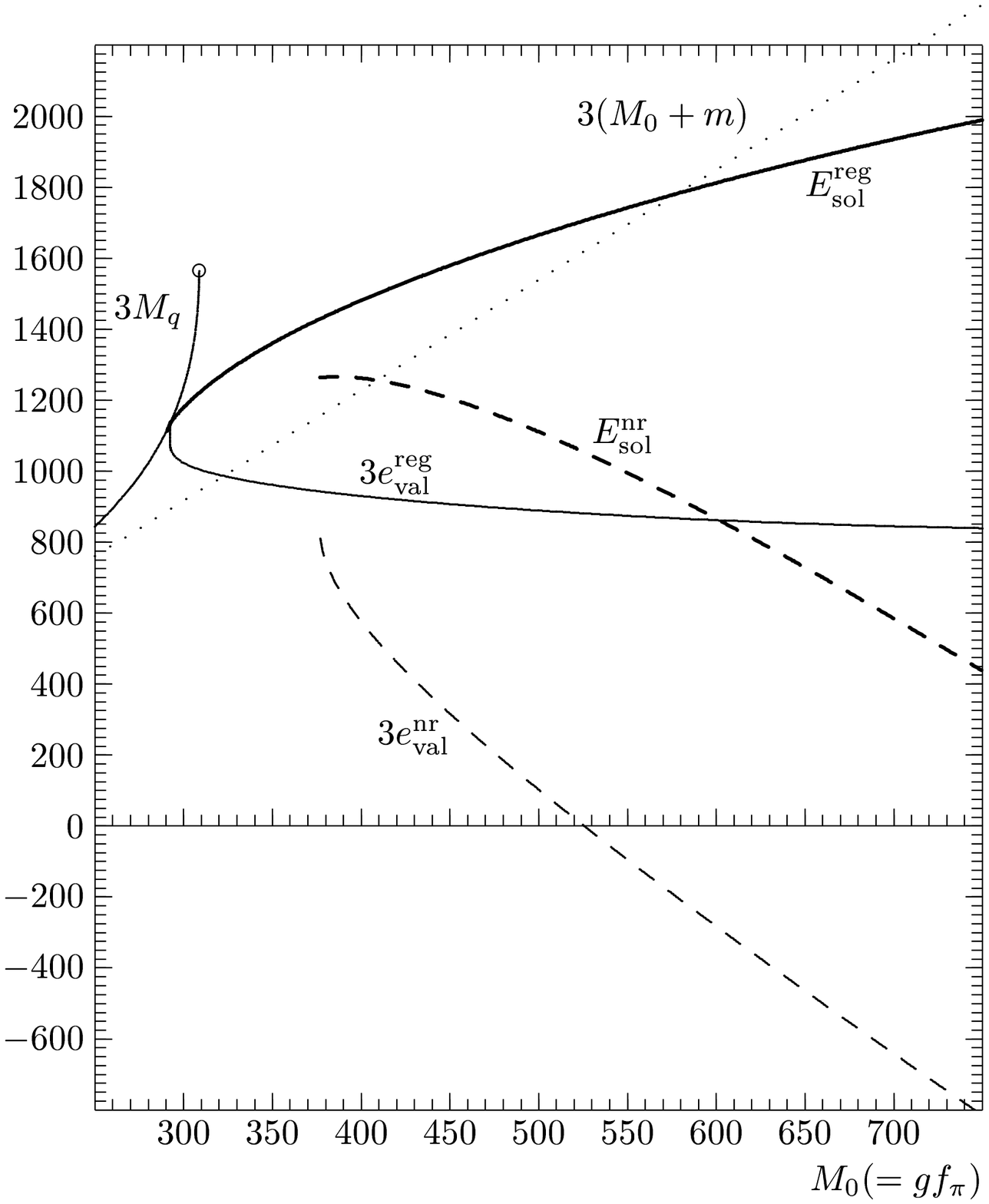}{90mm}
\caption{Solitons in linear $\sigma$-models: the
solid lines show the energy of the soliton (bold), the valence
contribution to the energy and of the mass of three free quarks 
plotted as functions of 
the parameter $M_0$  in the case when the valence state is regularized.
A 4-momentum Gaussian regulator is used; 
$\Lambda$ (see Figure~\ref{figep}) is fitted to $f_\pi=93$~MeV.
The dashed lines show the respective energies when
the valence orbit is not regularized and
the dotted line represents the corresponding stability line.
In this case quasi-stable soliton solutions can exist
which never happens in the regularized case.}
\label{figelsm}
\end{center}
\end{figure}

We assume that the Lagrangian (\ref{elsm-fam}) describes 
the sea quarks while the valence quarks are treated separately.
This is in the spirit of the approaches used in 
\cite{KRS,BB,GR,Mike,CB86}.
However, in all these approaches the valence orbit is not 
regularized and -- as we shall see in this section  -- this brings 
a qualitative difference with respect to the situation where it is
regularized.
If we agree that a 
regulator in the quark-quark or, equivalently, quark-chiral field 
interaction has a well grounded physical origin, 
the valence orbit should be regularized
in the same way as the orbits in the Dirac sea.

The valence orbit is determined as in (\ref{eval}).
The soliton energy then becomes (${\sigma'}=\sigma-f_\pi$)  :
$$
  E(\sigma,\pi_a) = 3e_\mathrm{val} + 
  \int \d^3\vec{r}\left(
\half\left[ (\nabla\sigma')^2 + m_\sigma^2{\sigma'}^2
          + (\nabla\pi_a)^2 + m_\pi^2\pi_a^2          \right]
  + g \left[ \sigma j^\sigma + \pi_a j_a^\pi \right]
                                                      \right)
  + E_{s.i.}\;,
$$
where the meson self-interaction term (the ``Mexican hat'') is given by
$$
  E_{s.i.}= {\lambda^2\over4}\,\int \d^3\vec{r}\left(
  {\sigma'}^4 + (\pi_a^2)^2 + 4f_\pi{\sigma'}^3 + 2{\sigma'}^2\pi_a^2
      + 4f_\pi{\sigma'}\pi_a^2 \right)\;.
$$
Here we have introduced the source terms which explicitly contain
the regulator (see (\ref{sceq})):
\begin{eqnarray}
  j^\sigma &=& N_q{\rm res}_v^{-1}\,
    q_{v_{\nu0}}^\dagger(\vec{r})\,\beta\,
  r^2((\nu_0^2-\vec{\nabla}^2)/\Lambda^2) q_{v_{\nu0}}(\vec{r})\;, 
\nonumber\\
  j^\pi_a &=& N_q{\rm res}_v^{-1}\,q_{v_{\nu0}}^\dagger(\vec{r})\,
   \i\beta\gamma_5\tau_a\,
  r^2((\nu_0^2-\vec{\nabla}^2)/\Lambda^2) q_{v_{\nu0}}(\vec{r})\;.
\label{source}
\end{eqnarray}
The mean-field solution is obtained by solving the self-consistency
equations (the hedgehog ansatz 
$\sigma(\vec{r}) = \sigma(r)$ and $\pi_a(\vec{r})= \hat{r}_a\pi(r)$
is assumed):
\begin{eqnarray}
&& \left({\d^2\over\d r^2} + {2\over r}\,{\d\over\d r}
  - m_\sigma^2\right){\sigma'(r)}
   = j^\sigma(r) +
   \lambda^2\left[({\sigma'(r)}+3f_\pi)\,{\sigma'(r)}^2 
    + ({\sigma'(r)} + f_\pi)\,\pi(r)^2\right]\;,
\nonumber\\
&&\left({\d^2\over\d r^2} + {2\over r}\,{\d\over\d r}
- {2\over r^2} - m_\pi^2\right)\pi(r) 
 = j^\pi_a(\vec{r})\hat{r}_a +
  \lambda^2\left[
  {\sigma'(r)}^2 + 2f_\pi{\sigma'(r)} + \pi(r)^2\right]\pi(r)\;.     
\label{sigmaeq}
\end{eqnarray}

\begin{figure}[t]
\begin{center}
\epsfg{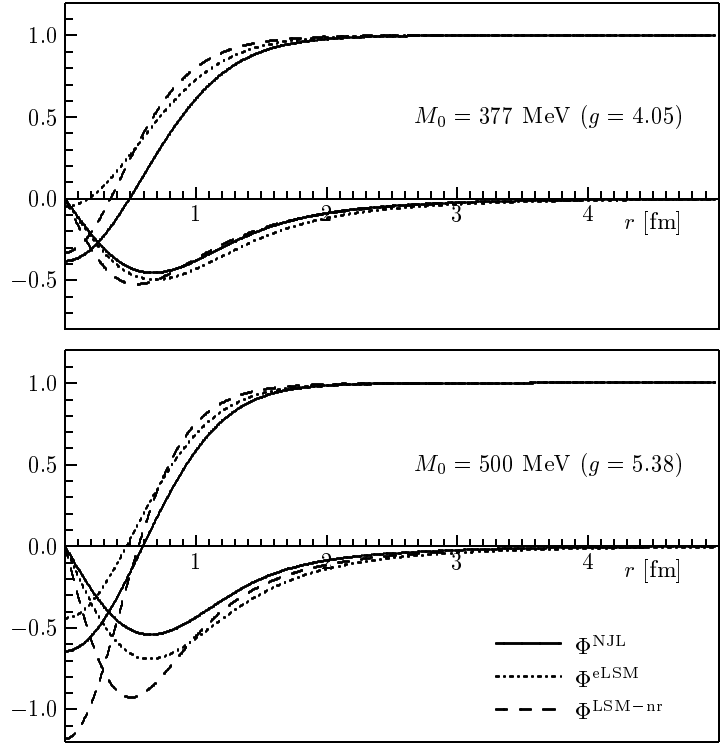}{95mm}
\caption{Comparison of the self-consistently determined fields 
in the NJL model ($\Phi^{\rm NJL}$)
and in two versions of the linear $\sigma$-model:
with ($\Phi^{\rm eLSM}$) and without ($\Phi^{\rm LSM-nr}$)
regularization of the  valence state.
A Gaussian regulator is used with $\Lambda$  fitted to 
$f_\pi=93$~MeV.
For $M_0=377$~MeV, the corresponding $\Lambda$ is relatively
large (578~MeV) and the three (pairs of) curves do not differ
considerably; for $M_0=500$~MeV, $\Lambda$ is much
lower (440~MeV) and the differences become more important:
the fields in the unregularized case acquire higher
gradients and the soliton shrinks.}
\label{mflsm}
\end{center}
\end{figure}

In Figures~\ref{figelsm} and \ref{mflsm} we compare 
the properties of the soliton  when 
the valence orbit is regularized and when it is not.
Close to the threshold the two solutions do not differ much 
-- as it should be -- 
since here the cut-off $\Lambda$ is relatively large.
For higher values of $M_0$ (or $g$) the behavior is qualitatively
quite different; in particular, the energy of the unregularized 
valence orbit soon becomes 
negative which never happens for the regularized one.

Let us finally mention why the regularization is important also 
in the linear $\sigma$-model with only valence quarks.
Even if there is no instability as in 
the regularized linear $\sigma$-model\footnote{%
In the regularized  model  both the meson and the quark degrees
of freedom are used in describing the Dirac sea which leads to
appearance of an unphysical pole for the $\sigma$-propagator
\cite{Ripka87,Perry87}.},
one encounters serious difficulties when going beyond the
mean field approximation.
In the $\sigma$-model it is possible to use the Peierls-Yoccoz 
projection of good linear momentum, spin and isospin 
in order to obtain physical nucleon states.
It turns out that the energy of the soliton after projection
is strongly reduced already when the mean-field solution is 
used for the chiral fields \cite{Neuber}.
If in addition one allows a variation of the chiral field 
profiles, one obtains a solution with an energy considerably
lower than the nucleon mass.
The energy gain is mostly due to a strongly localized
chiral field which lowers the valence energy.
Such a strong localization is not allowed when
the regulator is used since it puts a physical limit
on the gradients of the field (through the source
term (\ref{source})).


\begin{thebibliography}{99}

\bibitem{Goeke92}
P. Sieber, T. Meissner, F. Gr{\"{u}}mmer, and K. Goeke, Nucl. Phys. 
{\bf A 547} (1992) 459

\bibitem{Ripka93d}
 Th. Meissner, G. Ripka, R. W{\"{u}}nsch, P. Sieber, F. Gr{\"{u}}mmer, and K.
  Goeke, Phys. Lett. {\bf B 299} (1993) 183.

\bibitem{Ripka87}
G. Ripka and S. Kahana, Phys.\ Rev. {\bf D36} (1987) 1233.

\bibitem{Perry87} R. Perry, Phys.\ Lett. {\bf B199} (1987) 489.

\bibitem{Diakonov86}
D.~I. Diakonov and V.~Y. Petrov, Nucl. Phys. {\bf B 272} (1986) 457.

\bibitem{bled99}
W. Broniowski, INP Cracow preprint No.~1828/PH (1999), hep-ph/9909438, talk
  presented at the Mini-Workshop on {\em Hadrons as Solitons}, Bled, Slovenia,
  6-17 July 1999

\bibitem{GBR98}
B. Golli, W. Broniowski and G. Ripka,  Phys. Lett. {\bf B 437} 
(1998) 24.

\bibitem{Ripka84}
S. Kahana and G. Ripka, Nucl. Phys. {\bf A 429} (1984) 462.

\bibitem{Ripka97}
G. Ripka, Quarks Bound by Chiral Fields, Oxford University Press, 
Oxford, 1997.

\bibitem{KRS} 
 S. Kahana, G. Ripka and V. Soni, Nucl. Phys.{\bf A 415} (1984) 351.

\bibitem{BB} 
 M. C. Birse and M. K. Banerjee, Phys. Lett. {\bf B 136} (1984) 284; 
 Phys. Rev. {\bf D 31} (1985) 118.

\bibitem{GR} 
 B. Golli and M. Rosina, Phys. Lett. {\bf B 165} (1985) 347.

\bibitem{Mike} 
 M. C. Birse, Phys. Rev. {\bf D 33} (1986) 1934.

\bibitem{CB86}
T.~D. Cohen and W. Broniowski, Phys. Rev. {\bf D 34} (1986) 3472.

\bibitem{Neuber}
 T. Neuber and K. Goeke, Phys. Lett. {\bf B 281} (1992) 202.

\end{thebibliography}
\end{document}